\documentclass[10pt,article]{aastex}
\usepackage{emulateapj5}
\input{psfig.sty}
\catcode`\@=11 % This allows us to modify PLAIN macros.
\def\@versim#1#2{\vcenter{\offinterlineskip \ialign{$\m@th#1\hfil##\hfil$\crcr#2\crcr\sim\crcr } }}
\newcommand{\beq}{\begin{equation}}
\newcommand{\eeq}{\end{equation}}
\def\lsim{\mathrel{\mathpalette\@versim<}}
\def\gsim{\mathrel{\mathpalette\@versim>}}

\newenvironment{figurehere}
  {\def\@captype{figure}}
  {}
\makeatother

\begin{document}
\title{Synchrotron Radiation From Radiatively Inefficient Accretion Flow Simulations: Applications to Sgr A*} \author{Joshua E. Goldston\altaffilmark{1}, Eliot Quataert\altaffilmark{2}, and Igor V. Igumenshchev\altaffilmark{3}} \altaffiltext{1}{UC Berkeley, Astronomy Department, 601Campbell Hall, Berkeley, CA 94720; goldston@astron.berkeley.edu} \altaffiltext{2}{UC Berkeley, Astronomy Department, 601 Campbell Hall, Berkeley, CA 94720; eliot@astron.berkeley.edu} \altaffiltext{3}{Laboratory for Laser Energetics, University of Rochester, 250 East River Road, Rochester, NY 14623; iigu@lle.rochester.edu}

\medskip

\begin{abstract}

We calculate synchrotron radiation in three-dimensional
pseudo-Newtonian magnetohydrodynamic simulations of radiatively
inefficient accretion flows. We show that the emission is highly
variable at optically thin frequencies, with order of magnitude
variability on time-scales as short as the orbital period near the
last stable orbit; this emission is linearly polarized at the $\sim
20-50 \%$ level due to the coherent toroidal magnetic field in the
flow.  At optically thick frequencies, both the variability amplitude
and polarization fraction decrease significantly with decreasing
photon frequency.  We argue that these results are broadly consistent
with the observed properties of Sgr A* at the Galactic Center,
including the rapid infrared flaring.

\

\noindent {\it Subject Headings:} Accretion, accretion disks --
Galaxy: center

\end{abstract}

\section{Introduction}

The identification of magnetohydrodynamic (MHD) turbulence as a robust
angular momentum transport mechanism has led to a new era in the study
of accretion onto compact objects (see Balbus 2003 for a
review). Numerical simulations can now demonstrate accretion without
imposing an arbitrary anomalous viscosity. Although analytic steady
state models provide an important framework for understanding the
structure of accreting systems (e.g., Shakura and Sunyaev 1973;
Narayan and Yi 1994; Blandford and Begelman 1999), a computational
approach is required to capture the time-dependent, turbulent
evolution of the flow. Simulations open up the time domain to
theoretical study and should ultimately provide a more robust
description of the observed features of accreting systems.  Krolik and
Hawley (2002) exploited this possibility by computing power spectra of
quantities such as the accretion rate and Maxwell stress, both of
which are presumably related to the bolometric output of the accretion
flow. Hawley and Balbus (2002) carried out similar calculations,
including estimating the characteristic frequency for synchrotron
radiation in their simulations.  Here we extend these calculations by
explicitly calculating the radiative output from MHD simulations; for
reasons explained below, we focus on synchrotron radiation.  Our goal
is to make general connections between numerical simulations and
observations of accreting systems - we do not believe that the
simulations yet contain sufficient physics to warrant detailed
quantitative comparisons.  We focus on observables such as
polarization and fractional variability that are not as sensitive to
the absolute value of the predicted flux (which is difficult to
calculate for several reasons explained in \S2).
 
We apply our results to observations of Sgr A*, the $\approx 3.6
\times 10^6 M_{\sun}$ massive black hole (MBH) at the center of our
galaxy (Sch\"odel et al. 2002; Ghez et al. 2003). Sgr A* is known to
be remarkably faint: { the canonical Bondi accretion rate estimate
inferred using} the observed gas density around the black hole
predicts a bolometric luminosity $L \approx 0.1\dot{M} c^2\approx
10^{41}$ ergs s$^{-1}$, if radiation were produced with 10\%
efficiency (e.g., Baganoff et al. 2003). Instead, the observed
luminosity is $\sim 10^{36}$ ergs s$^{-1}$ (Melia and Falcke
2001). This disagreement favors radiatively inefficient accretion flow
(RIAF) models of Sgr A*, in which little of the gravitational binding
energy of the inflowing matter is radiated away (e.g., Narayan et
al. 1995; see Quataert 2003 for a review).  Numerical simulations of {
RIAFs have shown that part of the reason Sgr A* is so faint is that
the accretion rate is much less than the Bondi estimate (e.g., Stone
\& Pringle 2001; Hawley \& Balbus 2002; Igumenshchev et al. 2003).
This conclusion has been confirmed by observations of linear
polarization in the mm emission from Sgr A* (Aitken et al. 2000; Bower
et al. 2003; see, e.g., Agol 2000, Quataert \& Gruzinov 2000, and
Melia, Liu, \& Coker 2000 for how the observed polarization constrains
accretion models).}

Sgr A* has been detected from the radio to the X-rays, with most of
the bolometric power emerging at wavelengths near 1 millimeter (see, e.g.,
Melia \& Falcke 2001 for a review of spectral models). The millimeter
emission is significantly linearly polarized ($\sim 10\%$), consistent
with synchrotron radiation (Aitken et al. 2000; Bower et al. 2003),
while lower frequency radio emission is circularly polarized at the
$\approx 0.1-1 \%$ level (Bower et al. 1999). For the purposes of this
paper, the most interesting aspect of the observations is the strong
variability detected across the entire spectrum. X-ray flares with
amplitudes from factors of a few up to $\sim50$ have been detected on
hour time-scales (Baganoff et al. 2001; Porquet et al. 2003). Flaring
has also been seen in the infrared on time-scales as short as 10s of
minutes (Genzel et al. 2003; Ghez et al. 2004). Order unity
variability in the millimeter has been detected on longer time-scales
(week to month; Zhao et al. 2003), with even lower amplitude, longer
time-scale variations in the radio (Herrnstein et
al. 2004). Occasionally, the radio emission can vary more rapidly as
well, by $\sim 25\%$ over a few hours (Bower et al. 2002).  In
addition to variability in the total flux, the polarization in the
radio exhibits variability as well (Bower et al. 2002, 2005). The
origin of these variations is not fully understood.  Analytic models
generally interpret the X-ray and IR flares as emission due to
transient particle acceleration close to the black hole (e.g., Markoff
et al. 2001; Yuan, Quataert, \& Narayan 2003, 2004; hereafter YQN03
and YQN04), although alternative interpretations exist (e.g., Liu \&
Melia 2002; Nayakshin et al. 2004).
 
In this paper we compute synchrotron radiation in an MHD simulation of
a RIAF and discuss the results in the context of observations of Sgr
A*. We focus our analysis on synchrotron radiation because it probably
dominates the bolometric output of RIAFs at very low accretion rates
(see, e.g., Fig. 5 of YQN04), and because of its clear applicability
to Sgr A* (at least in the radio-IR).  The synchrotron radiation is
calculated ``after the fact,'' using the results of existing numerical
simulations that do not incorporate radiation.  This is
self-consistent because energy loss via radiation is negligible and
does not modify the basic dynamics or structure of the accretion flow.

The plan for the rest of this paper is as follows.  In the next
section (\S2) we describe the simulations and our method for
calculating synchrotron radiation. %; we also point out some of the
%important uncertainties in our analysis.  
We then describe our results in \S3.  We begin by considering
optically thin synchrotron emission (both total intensity and
polarized), and then consider optical depth effects.  Finally, in \S4
we conclude and discuss our results in the context of observations of
Sgr A*.  Before describing our method in detail, it is worth noting
some of the simplifications in our analysis.  The simulations used in
this paper are Newtonian and so do not account for relativistic
effects near the black hole.  In keeping with this simplified
dynamics, we do not account for Doppler shifts or gravitational
redshift when calculating the radiation from the flow.  Lastly, the
MHD simulations used here do not have detailed information about the
electron distribution function, which we parameterize as a Maxwellian
with a power-law tail.
 
\section{Synchrotron Radiation from MHD Simulations}

 The specific simulation used in this paper is `Model A' described in
Igumenshchev et al. (2003).  This is a Newtonian MHD simulation using
the Paczynski-Wiita potential.  The domain of the simulation is a set
of 5 nested Cartesian grids, each of which has 32 elements in the x
and y directions and 64 elements in the z direction. The smallest grid
has elements that are 0.5 Schwarzschild radii ($R_s$) across, where
$R_s = 2G{M}/c^2$, and each successively larger grid has elements
twice as large as the previous grid. All of the grids cover only two
octants, both positive and negative z, but only positive x and y.  The
boundary conditions are periodic such that the x = 0 plane is mapped
onto the y = 0 plane.  At small radii the black hole is represented by
an absorbing sphere and the central 2 $R_s$ are not evolved.  At the
outer boundary, $R = 256 R_s$, outflow boundary conditions are
applied.  Matter is injected continuously near the outer boundary with
toroidal magnetic fields, allowing a steady state to be reached in
which inflow into the grid is balanced by outflow out of the
simulation domain and into the ``black hole.''  The flow has
zero net poloidal magnetic field, and forms a geometrically thick
convective accretion disk.  There are no jets or significant mass
outflow in the polar direction.
 
Since the simulation variables are dimensionless, length-scales and
time-scales can be adjusted for any black hole mass.  We present all
physical quantities using $M = 3.6 \times 10^6 M_\odot$, appropriate
for Sgr A*.  The data we use are taken from the roughly steady state
portion of the simulation.  We work with two different data sets. Data
set 1 consists of 32 time-slices staggered by roughly 30 hours, which
is the orbital period at about 50 $R_s$. These data include the full
3D structure of the flow (density, pressure, magnetic field strength,
etc.).  Since much of the emission comes from close to the BH,
however, there is significant evolution between these
time-slices. Data set 2 thus consists of about 10,000 time-slices
separated by about $30$ sec, which corresponds to about 1/2 of the
light-crossing time of the black hole's horizon.  {  These data were
generated during a run of the simulation, and only contain information
about the synchrotron radiation not the structure of the flow. Because
of memory constraints, we could not retain the full 3D structure of
the flow at all 10,000 of these timesteps.  Instead, we rely on data
set 1 (which has the full 3D structure, but sparser time sampling) to
correlate the synchrotron emission with the structure of the flow and
to determine optical depth effects.} We use data set 2 to probe the
optically thin emission and variability on all relevant time-scales.

There are several significant difficulties in using the simulations to predict synchrotron radiation.  First, the density of gas in the simulation is arbitrary because the accretion rate is not fixed in physical units; that is, although the simulation describes the spatial and temporal evolution of the density, a single normalization parameter must be specified to compute physical gas densities. We adjust this density normalization so that the total synchrotron power is roughly that observed from Sgr A* (this depends on electron temperature; see below).  Given the density normalization, the total pressure and magnetic field strength are uniquely determined.

The biggest uncertainty in our analysis is that the simulations only
evolve the total pressure, while synchrotron radiation is produced by
the electrons.  Since RIAFs are collisionless plasmas, there is no
reason to expect equal electron and  ion temperatures or a thermal
distribution of electrons (e.g., Quataert 2003). This is problematic because the synchrotron emission is sensitive to the exact electron distribution function, about which we have no information from MHD simulations.  We account for this uncertainty as best as we can: we  model the electrons using a thermal distribution with a non-thermal  power-law tail.  Given a total ion + electron temperature (pressure)  $T_{tot} = T_i + T_e$ from the simulations, we determine the electron temperature using $ T_e/T_{tot}$ as a free parameter of our analysis. We show results for several values of $T_e/T_{tot}$, focusing on $T_e/T_{tot} = 1$ and $T_e/T_{tot} = 1/4$.  This roughly brackets the range of electron temperatures in recent semi-analytic models of emission from Sgr A* (e.g., YQN03). It is important to stress again that there is no compelling reason to expect the electron temperature to be everywhere proportional to the total temperature, but this is the best that we can do without solving separate ion and electron energy equations or carrying out kinetic simulations.

The choice of density normalization and electron temperature
parameterization have a nontrivial impact on the predicted spectrum of
the accretion flow.  Figure \ref{plot:ONE} shows the optically thin
synchrotron emission from a uniform plasma with $T_{tot}=2
\times10^{11}$ K for different choices of $T_e/T_{tot}$.  For each
calculation, we have adjusted the density to fix the total synchrotron
power.  To understand these results analytically, note that the peak
frequency for relativistic thermal synchrotron emission (where $\gamma
\propto T_e$) scales as \beq \nu_{peak} \propto B \gamma^2 \propto
q^{\frac{1}{2}}\left({T_e/T_{tot}}\right)^2 \eeq and the total
synchrotron power scales as \beq P_{total} \propto B^2 \gamma^2 n_e
\propto q^2\left(T_e/T_{tot}\right)^2 \propto q^{3/2} \nu_{peak} \eeq
where $q$ is the density normalization parameter (i.e., $n_e \propto
q$) and we assume fixed $\beta = P/(B^2/8\pi)$.
 
Given these uncertainties, our strategy is to choose several
parameterizations for the electron temperature (via $T_e/T_{tot}$) and
to adjust the density normalization such that the total synchrotron
power is reasonable.  Physically, this corresponds to electron
temperatures $\sim 10^{10}- 10^{11}$ K near the black hole and gas
densities $\sim 10^6-10^7$ cm$^{-3}$ (see also the one-dimensional
analytic models in YQN03).\footnote {For the parameters chosen here,
neglecting radiation in the dynamics of the accretion flow is an
excellent approximation.  Moreover, the synchrotron cooling time for
the bulk of the electrons (those with $\gamma \sim kT_e/m_e c^2$) is
usually much longer than the inflow time, so that the electron
temperature is determined by a balance between heating and advection,
rather than heating and cooling. As a result, the only way to model
the electron energetics better than we have here is to calculate a
separate electron energy equation.  Were the electron cooling time
very short (appropriate at much higher densities than we consider
here), one could calculate the electron temperature everywhere in the
accretion flow by locally balancing heating and cooling, without
worrying about the Lagrangian derivative of the electron internal
energy.}  Because the simulations cannot uniquely predict the total
synchrotron power or peak synchrotron frequency, our analysis focuses
on observables that are less sensitive to the overall normalization of
the emission, namely the variability and polarization.  As we show
later, the semi-quantitative conclusions of this paper do not depend
that sensitively on the uncertainties in $T_e/T_{tot}$ and $q$, though
detailed quantitative comparisons between simulations and observations
are clearly premature.
 
In addition to considering a thermal electron population, we also
account for a non-thermal tail in the electron distribution
function. This power-law component is strongly motivated by models for
the IR and X-ray emission from Sgr A* (e.g., Markoff et al. 2001;
YQN04), and also is expected theoretically in models of collisionless
plasmas.  We assume that the power-law tail has a fraction $\eta$ of
the electron thermal energy and a power-law index $p$ where $n(\gamma)
\propto \gamma^{-p}$.  We fix $p = 3$ and choose either $\eta = 0$ (no
power-law component) or $\eta = 0.05$ (see YQN03).  We do not vary $p$
or $\eta$ in time or space, although such variation is likely inevitable due to transient events such as shocks and reconnection (and is certainly suggested by the IR and X-ray flaring from Sgr A*).  As a result, our calculations of variability at high frequencies are likely {\it lower limits}, since they do not include variations in the accelerated electron population.

We compute the total synchrotron emission of data sets 1 and 2
assuming ultrarelativistic electrons (following Pacholczyk 1970,
henceforth P70).  We have checked that using the transrelativistic
fitting formula of Mahadevan et al. (1996) yields almost identical
results to the ultrarelativistic calculation for our problem.  Given
uncertainties in our viewing angle towards the accretion flow, we use
a simple angle averaged emissivity to compute the emission from data
set 2 (the high time sampled data set).  We have assessed the validity
of this approximation by computing the angle-dependent emission from
data set 1 (which has the full 3D structure): the true flux is
typically similar to that given by the angle-averaged formula to
within $\approx 50 \%$ (much less than the theoretical uncertainties
due to the electron distribution function).  The disagreement can be
somewhat worse at very high frequencies ($\nu \gsim 10^{14}$ Hz) when
one is on the exponential tail of the synchrotron emissivity.

In addition to calculating the total flux, we also calculate the
linear polarization of synchrotron radiation in the simulations, again
assuming ultrarelativistic electrons (P70; eqn 3.39).  We also present
several calculations including the effects of synchrotron
self-absorption. For simplicity, we only calculate self absorption for
viewers in the equatorial plane of the accretion flow. For our typical
parameters ($T_e/T_{tot}$ and $q$), self-absorption effects become
important below frequencies of a few hundred GHz.  We account for
self-absorption in the thermal component by calculating the optical
depth along a ray using the emission coefficient and the thermal
source function $S_{\nu} = 2kT_e{\nu}^2/c^2$.  The intensity along
each ray is then given by adding up the optically thin emission from
each grid point along the ray, weighted by $\exp(-\tau)$.  Note that
power-law electrons are omitted from this calculation, but each
polarization mode is calculated separately so we can determine the
polarization at both optically thick and thin frequencies.

We assume that at each time the observed emission is given by the {\it steady state} solution to the radiative transfer equation given the temperature, density, etc. of the accretion flow at that time. This is likely to be a reasonable approximation except for very short time-scale variability, and for emission from radii where the velocity of the gas approaches the speed of light (these limits are, of course, of considerable interest, but relaxing the steady state assumption is a significant effort beyond the scope of this work). In keeping with the Newtonian dynamics in the simulations used here, we also do not include gravitational redshift or Doppler shifts.

It is important to point out the effects of finite spatial resolution
on our calculations.  High frequency thermal synchrotron emission is
produced primarily at small radii close to the black hole, where the
temperature and magnetic field strength are the largest.  The spatial
resolution in this region is $0.5 R_s$, so that at the last stable
orbit ($3 R_s$), there are a rather small number of grid points.  We
find that the thermal synchrotron flux at high frequencies (at and
above the thermal peak) and the location of the peak are often
determined by emission from a small number of grid points.  As a
result, the spatial resolution is not quite sufficient to accurately
calculate the highest frequency synchrotron emission.  This is likely
to be a generic problem in calculating emission from simulations given
resolution constraints.  It is particularly acute above the thermal
peak because the emission there is exponentially sensitive to $B$ and
$T_e$.  By contrast, below the thermal peak the emission is always
well determined because it comes from a larger range of radii.  As a
check on how sensitive our results are to the inner few Schwarzschild
radii, we have assessed how the emission from the flow changes if we
only consider emission from outside (say) 4 or 6 $R_s$.  We find that,
although the overall amplitude of the flux at very high frequencies
decreases, our primary conclusions regarding fractional variability
and polarization (as enumerated in \S4) are essentially unchanged.
Thus we believe that these results accurately reflect the underlying
dynamics of the accretion flow.

% if we only consider emission from outside (say) 4 or 6
%$R_s$.  In addition, we find that our results are very similar below
%and above the thermal peak (see Fig. 5).  
%Thus we believe that the y are not that sensitive to
%spatial resolution and are instead characteristic of the accretion
%flow dynamics.
 
\section{Results}

To begin we present results for unpolarized synchrotron emission,
without taking into account the effects of synchrotron self-absorption.  We then discuss the polarization of the emission and the effects of synchrotron self-absorption.  For some of the frequencies considered below it is not self-consistent to ignore self-absorption, but we do so initially to illustrate general points about synchrotron emission from the flow.

\subsection{Optically Thin Emission}

\begin{figure}
\begin{center}
\plotone{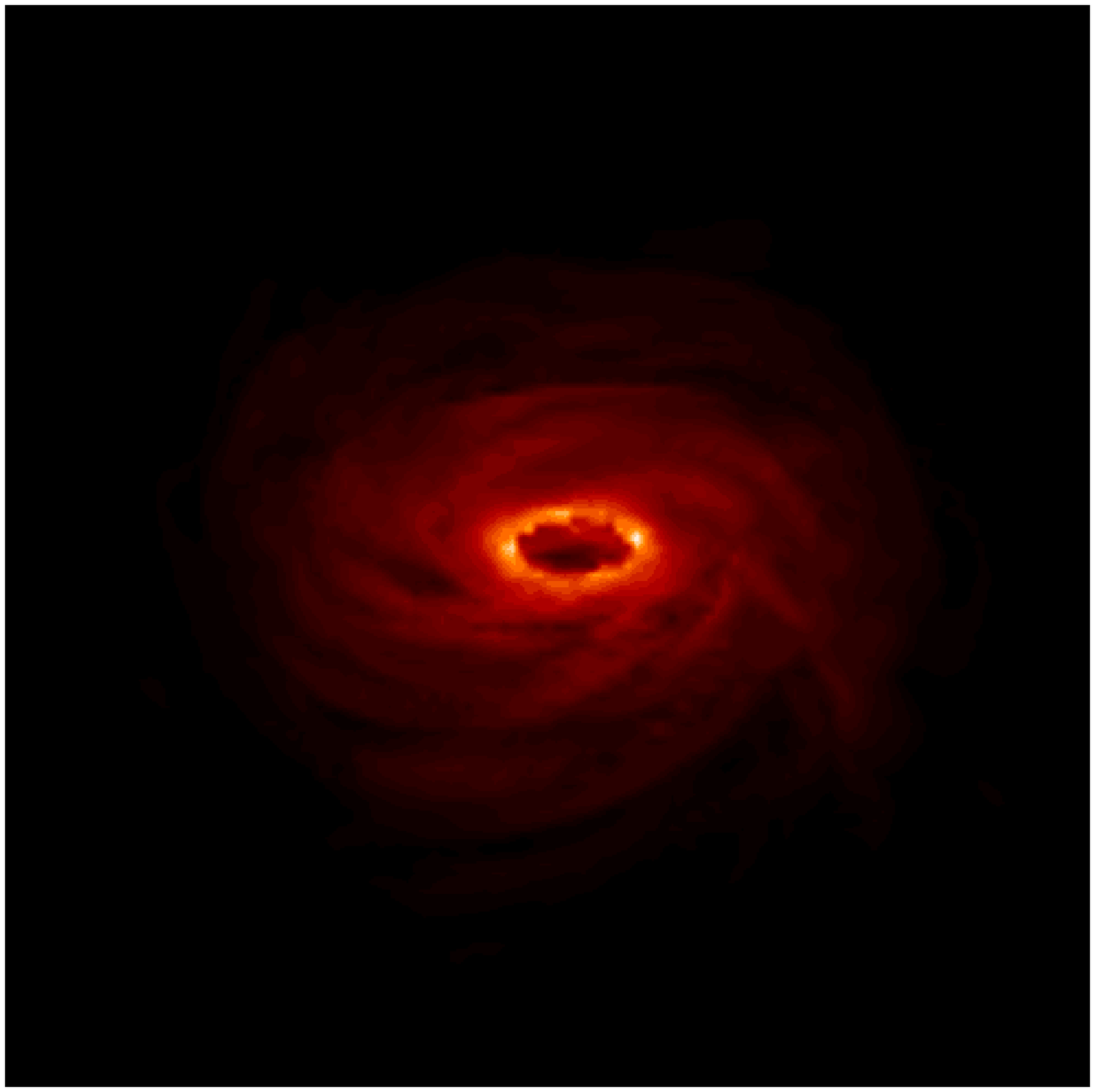}
\end{center}
\caption{An image of the central 16 $R_s$ at a particular time-slice for $\nu \approx 450$ GHz, near the peak of the thermal emission. The intensity scale is logarithmic and the brightest area is in the equatorial plane near the last stable orbit, where the density, temperature, and magnetic field strength are the largest.  For $\nu \ll 450$ GHz (below the thermal peak), the emission is much less centrally concentrated.}
\label{plot:ZERO}
\end{figure}

To become familiar with the emission it is useful to examine a
time-slice and `see' what the flow would look like if it could be
resolved with an array in the sub-mm.  This is shown in Figure
\ref{plot:ZERO} for a typical time-slice, as viewed from an arbitrary
angle off of the plane at $\nu \approx 450$ GHz, which is near the
peak in the thermal emission (see Fig. \ref{plot:EIGHT}). We only show
the central 16 $R_s$, which is by far the dominant emission region
(the intensity plot in Fig. \ref{plot:ZERO} is logarithmic). Most of
the emission originates close to the central hole, in the equatorial
plane, rather than, for instance, in polar outflows or jets, which do
not appear in this simulation.  This structure changes somewhat from
time-slice to time-slice but the prominence of an equatorial region
near the hole is common to all. At lower frequencies, where $\nu \ll
\nu_{peak}$, large radii are more important and so the emission is
much less centrally concentrated.  Note that the calculations shown in
Figure \ref{plot:ZERO} do not include any General Relativistic photon
transport (or even obscuration by the BH), which may give rise to
unique signatures in the observed image (e.g. Falcke, Melia, \& Agol
2000).  Nonetheless, these calculations are encouraging because the
emission arises so close to the hole, where GR effects would indeed be
important.
 
\begin{figure}
%\begin{center}
%\includegraphics[height=6.6in, width=5.1in, angle=90]{FIGURETWO}
\hspace*{1.2cm}
\centerline{
\psfig{file=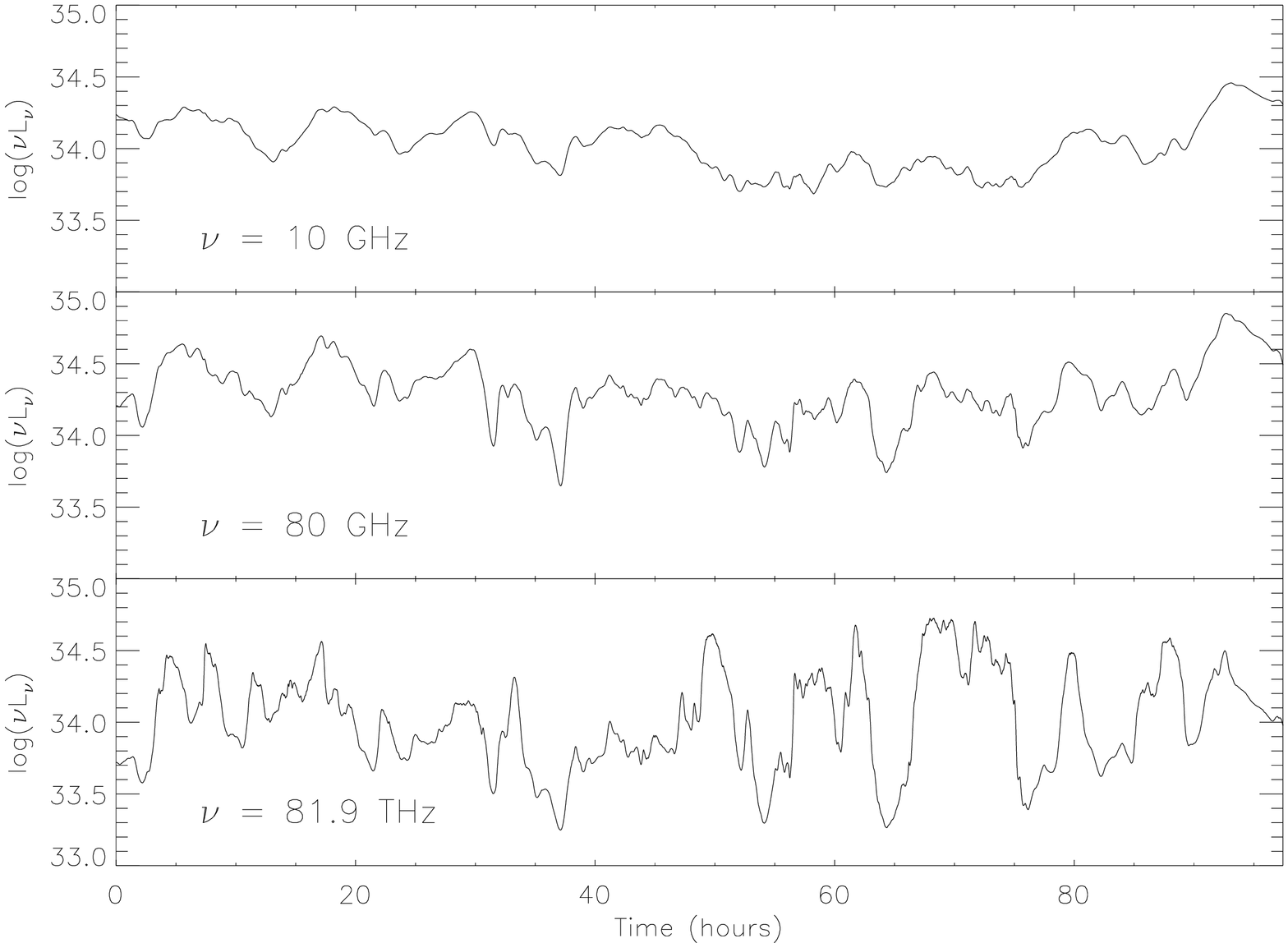,angle=90,width=7.5in}}
%\plotone{FIGURETWO.eps}
%\end{center}
\caption{Optically thin light-curves at three representative
frequencies, for $T_e/T_{tot} = 1/4$ and $\eta = 0.05$, where $\eta$
is the fraction of the electron thermal energy in a power-law tail
with $n(\gamma) \propto \gamma^{-3}$. The emission at $\nu = 10$ and
$80$ GHz is produced by the thermal electrons, while the $\nu = 81.9$
THz emission (K-band IR) is produced by the power-law electrons.}
\label{plot:TWO}
\end{figure}

Figure \ref{plot:TWO} shows the variability of the emission with time
at three representative frequencies, from the radio to the IR. The
calculations are for $T_e/T_{tot} = 1/4 $ and a power-law tail with
$\eta = 0.05$.  The $\nu = 10$ GHz and $\nu = 80$ GHz emission are
produced by the thermal component, while the $\nu = 81.9$ THz emission
(K-band IR) is produced by the non-thermal tail.  The most striking
feature of these results is the strong variability of the emission
with time. The flux varies by up to an order of magnitude, often on
time-scales as short as an hour. Figure \ref{plot:TWO} also shows that
the amplitude of the variability increases, and the time-scale for
variability decreases, at higher photon frequencies.  This is because
higher photon frequency emission originates closer to the hole where
the dynamical time-scales are shorter. In addition, we find that
variability of the flow parameters in the inner region is larger than
the variability at large radii, accounting for the fact that the
higher frequency emission that originates close to the hole is more
variable.
 
\begin{figure}
%\begin{center}
%\includegraphics[height=6.6in, width=5.1in, angle=90]{FIGURETHREE}
%\plotone{FIGURETHREE.eps}
%\end{center}
\hspace*{-.2cm}
\centerline{
\psfig{file=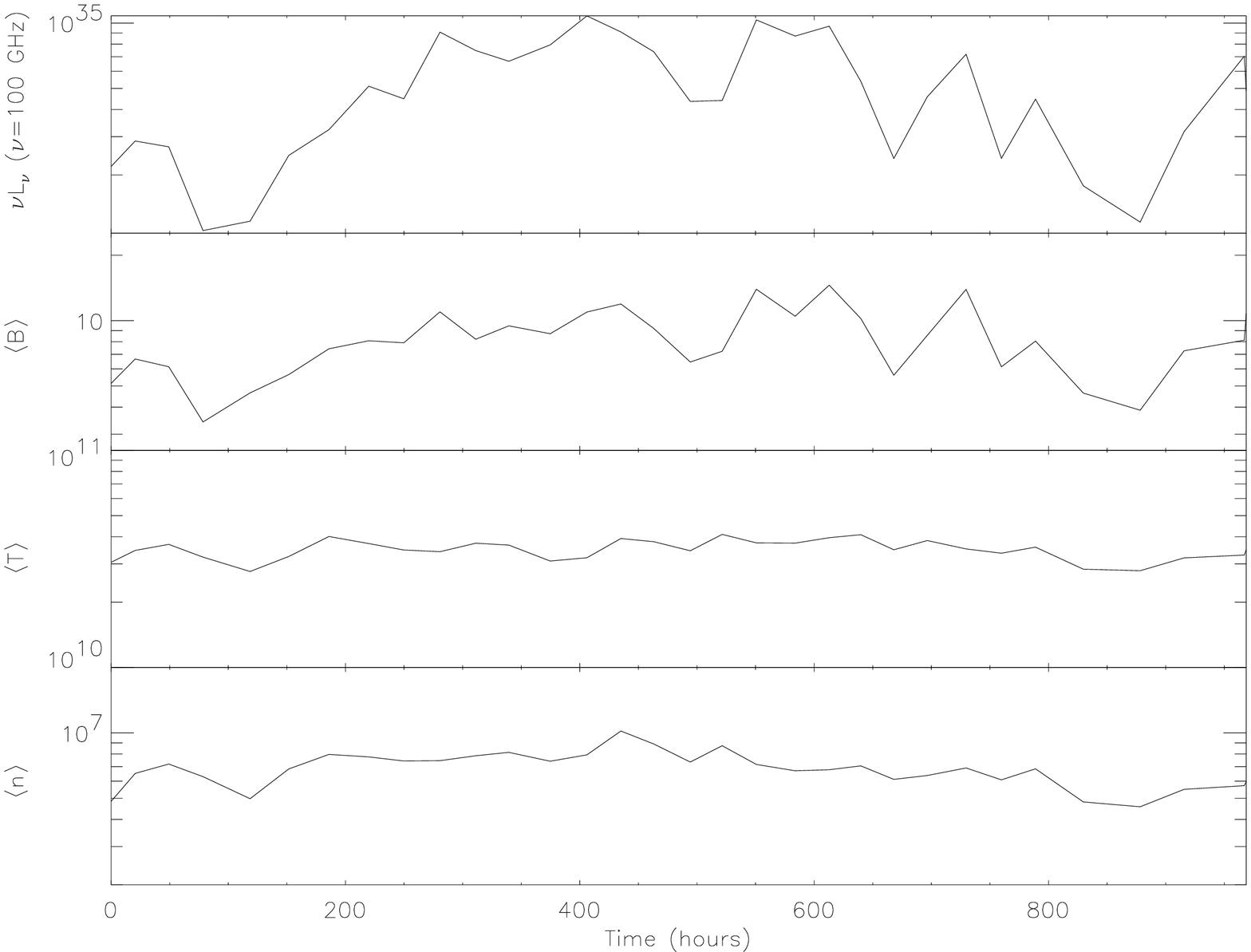,angle=90,width=7.3in}}
\caption{A light-curve at 100 GHz (top) plotted against the mean values of $B$, $n$ and $T_e$ within 6 $R_s$. All values are plotted from the $T_e/T_{tot} = 1/4$ model.}
\label{plot:THREE}
\end{figure}

To demonstrate the origin of this variability, Figure \ref{plot:THREE}
shows how the flux at 100 GHz varies on long time-scales, as compared
with the magnetic field strength ($|B|$), temperature ($T_e$), and
density ($n_e$), averaged from 2 to 6 $R_s$.  It is clear that the
variability is primarily due to changes in the magnetic field
strength, with fluctuations in temperature and density being less
important { (the correlation coefficients between $\nu L_\nu$ and $B$,
$n$, and $T$, are 0.93, 0.66, and 0.57, respectively).}  Note that
because the flow has $\beta \sim 10-100$, even large changes in the
magnetic field strength do not necessarily lead to appreciable
pressure or density changes, though they do cause significant
variability in the synchrotron emission.
 
\begin{figure}
\begin{center}
%\includegraphics[height=6.6in, width=5.1in, angle=90]{HOURPOWER}
%\plotone{HOURPOWER.eps}
\hspace*{-.4cm}
\centerline{
\psfig{file=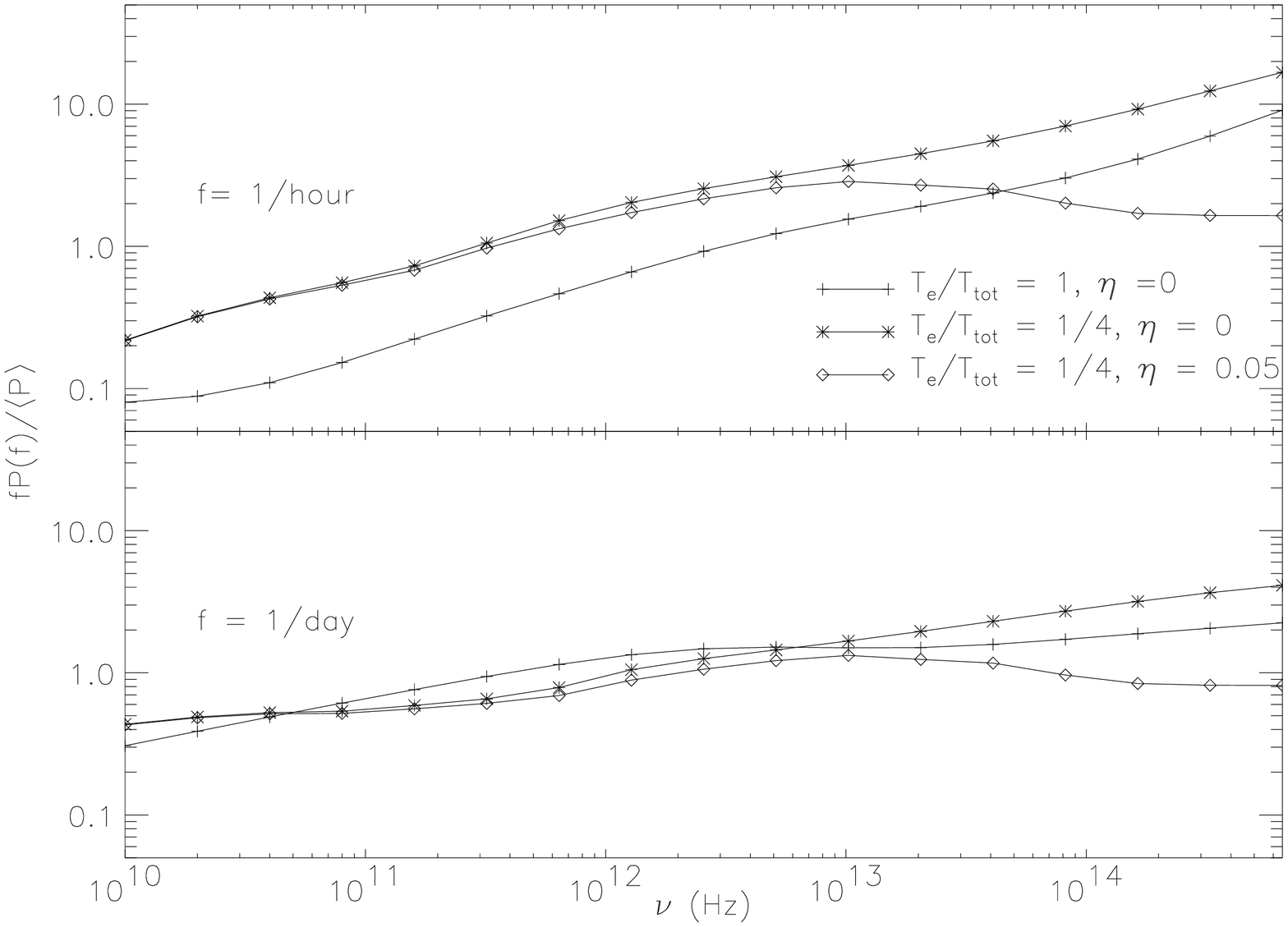,angle=90,width=7.3in}}
\end{center}
\caption{Each panel displays variability for 3 different models of the
electron thermodynamics as a function of photon frequency ($\eta$ is
the fraction of the electron energy in a power-law tail). The top
panel shows hour time-scale variability while the bottom is for day
time-scale variability. The Y-axis is a measure of the normalized
variability on a particular time-scale, where $\langle P\rangle$ is
the time average of $\nu L _{\nu}$ and $P\left(f\right) =
|\mathcal{F}\left(\nu L_{\nu}({\rm t})\right)|$, where $\mathcal{F}$
is the Fourier transform.  These calculations assume optically thin
synchrotron emission; optical depth effects are considered in
Fig. \ref{plot:EIGHT}.}
\label{plot:hrdy}
\end{figure}

In the previous section we emphasized that the predicted emission depends on how we parameterize the electron temperature and distribution function. Figure \ref{plot:hrdy} contrasts these different parameterizations in terms of the variability of the flow on different time-scales and at different photon frequencies.  Figure \ref{plot:hrdy} displays the power in Fourier components corresponding to one day and one hour periods, normalized to the mean flux (see caption for details).  On hour time-scales the two models with purely thermal electrons look very similar, except that the variability at a given frequency is smaller for the model with hotter electrons. The reason for this is that higher electron temperatures imply that at a given photon frequency the emission arises from larger radii (so that $B$ is smaller), where the dynamical timescales are longer and thus the short time-scale ($\sim$ hour) variability is weaker.
 
Figure \ref{plot:hrdy} also shows that higher photon frequency
emission is significantly more variable than lower photon frequency
emission on hour time-scales.  This is because the lower photon
frequency emission includes significant contributions from large radii
where the orbital time-scales that govern the variability of the flow
are $\gg 1$ hour.  The trend of increasing variability with increasing
photon frequency is present, but noticeably less dramatic, on day
time-scales.  This is simply because most of the radii that contribute
to the observed emission have orbital periods $\lsim 1$ day, and thus
are significantly variable on $\sim$ day time-scales. Finally, Figure
\ref{plot:hrdy} shows that non-thermal emission begins to dominate
thermal emission at $\nu \approx 10^{12}-10^{13}$ Hz.  In addition,
the models with pure thermal emission are more variable than those
with non-thermal emission at these frequencies; this is because $\nu
\gsim 10^{12}-10^{13}$ Hz is near or above the peak in the thermal
synchrotron emission (see Fig. \ref{plot:EIGHT}), and so the flux is
exponentially sensitive to variations in the flow parameters. Of
course, the flux (and the variable flux) in the non-thermal tail at
high frequencies is much larger than that produced by the thermal
electrons, but the {\it normalized} variability is smaller.
 
%At low photon frequencies there is more contribution from the outer
%radii with characteristic frequencies nearer to 1 day, and at high
%photon frequencies, where the dominant contribution is from the inner
%radii, there is less variability on the day time-scale.

\subsection{Linear Polarization}

\begin{figure}
%\begin{center}
%\includegraphics[height=6.6in, width=5.1in, angle=90]{FIGURESIX}
\hspace*{-.1cm}
\centerline{
\psfig{file=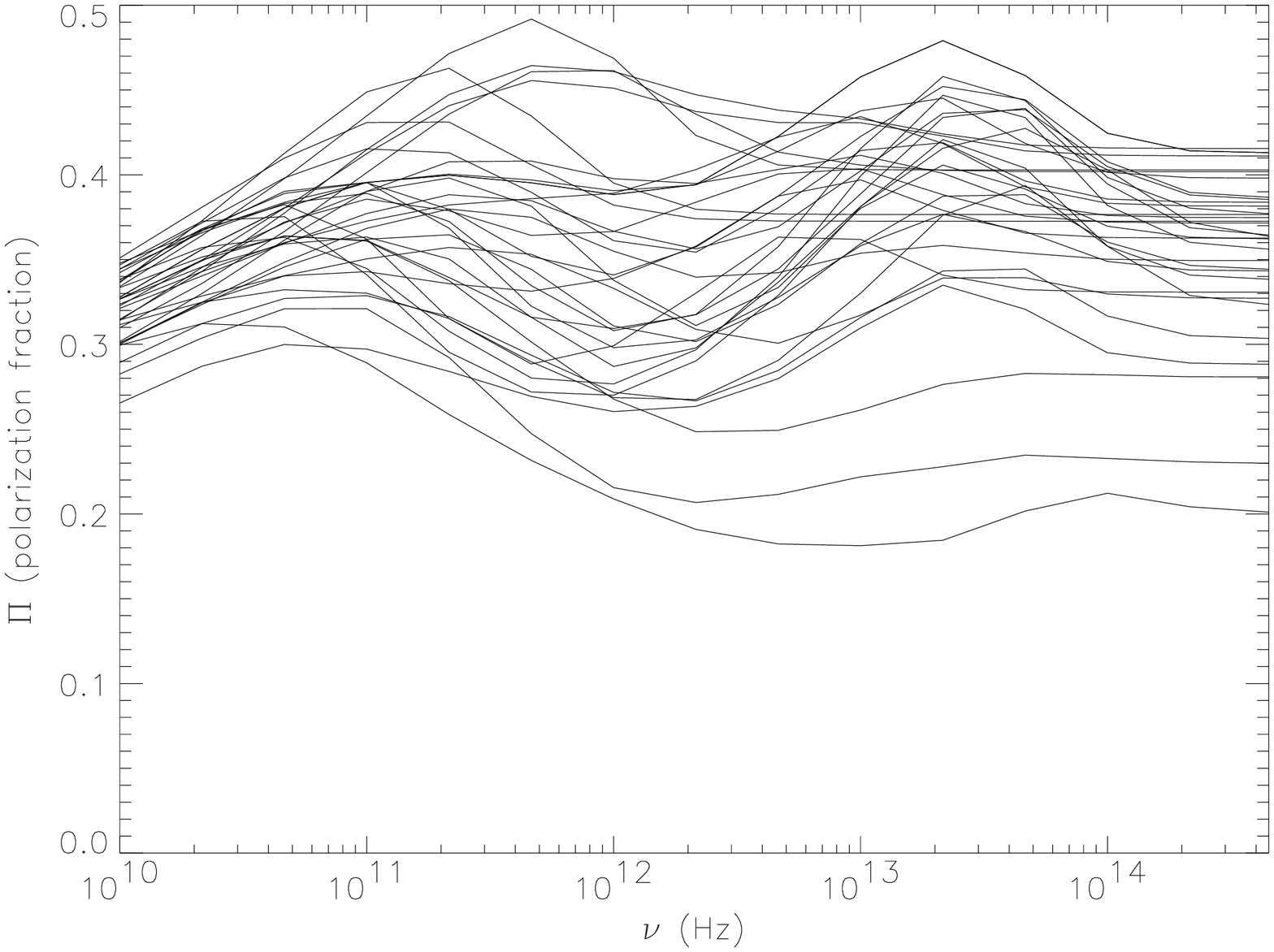,angle=90,width=8in}}
%\plotone{FIGURESIX.eps}
%\end{center}
\caption{Optically thin linear polarization fraction as a function of
frequency for each of the 32 time-slices for which we have the full 3D
structure (see Fig. \ref{plot:NINE} for optical depth effects). Each
time-slice is separated by $\approx$ 30 hours.  All curves assume
$T_e/T_{tot} = 1/4$ and $\eta = 0.05$.}
\label{plot:SIX}
\end{figure}

We have also computed the net linear polarization of synchrotron
emission in the simulation (neglecting Faraday rotation and optical
depth effects; the latter are considered in the next section).  The
results are shown as a function of frequency in Figure \ref{plot:SIX}
for all of the time-slices in data set 1 (each of which is separated
by about 30 hours); for these calculations we chose to view the flow
30 degrees off of the equatorial plane, set $T_e/T_{tot} = 1/4$, and
included a non-thermal tail with $\eta = 0.05$.  As Figure
\ref{plot:SIX} shows, the magnitude of the linear polarization is
substantial, typically $30-40 \%$.  There is some variation with
frequency, though it is not particularly dramatic.  The polarization
vector lies perpendicular to the equatorial plane.  It arises due to 
the coherent
toroidal magnetic field in the flow, analogous to the polarization of
synchrotron radiation in the Galaxy (e.g., Beck 2001). The
polarization would be larger ($\approx 60\% $; see
Fig. \ref{plot:NINE}) if we viewed the flow edge-on, while it would
vanish for face-on viewing angles.  Note also that there is
appreciable variability in the magnitude of the polarization, roughly
order unity changes on day to week time-scales. The variability is
somewhat larger at high frequencies, again because this emission
arises closer to the black hole.

\subsection{Synchrotron Self-Absorption}

In the previous sections we have assumed that the synchrotron emission
is optically thin at all frequencies of interest.  This assumption is
incorrect at low frequencies where synchrotron self-absorption becomes
important and the emission becomes optically thick.  In this section
we present initial results on the emission and variability at
optically thick frequencies, leaving a more comprehensive discussion
to future work.  We have computed the total flux as a function of
frequency including self-absorption for the case when the flow is
viewed in the equatorial plane, perpendicular to the rotation axis
(see \S2 for the method); we considered thermal electrons only and
took $T_e/T_{tot} = 1/4$.  The top panel in Figure \ref{plot:EIGHT}
shows the average and RMS emission both with (solid) and without
(dashed) optical depth effects, while the bottom panel shows the
normalized RMS variability for these two cases.\footnote{To check
whether the optically thick results at low frequency are sensitive to
spatial resolution (e.g., resolving the $\tau \sim 1$ surface), we
interpolated the flow structure to a finer grid and recalculated the
emission.  The results were essentially identical to those in Figure 7
for $\nu \lsim 10^{12}-10^{13}$ Hz.  The higher frequency (optically
thin) emission did change somewhat, however, because it is dominated
by very small radii (see the discussion at the end of \S2).}  The
averages and RMS are taken over all 32 time-slices (data set 1) for
which we have the full 3D structure necessary to compute optically
thick emission. Since these time-slices are separated by $\sim$ 1 day,
we do not have good constraints on short time-scale variability for
the optically thick emission. Note also that the RMS variability
plotted in Figure \ref{plot:EIGHT} yields a quantitative measure of
variability that is systematically smaller (by a factor of few) than
the Fourier analysis shown in Figure 5 (compare with the f = 1/day
plot), despite the fact that the data are essentially the same.  The
Fourier amplitude is a better measure of the `by-eye' variability, but
with only 32 (somewhat unevenly distributed) time-slices a
Fourier-amplitude version of Figure 7 was not feasible.

%such that the results in Figure
%\ref{plot:EIGHT}b and \ref{plot:hrdy} are in rough accord.

Figure \ref{plot:EIGHT} shows that self-absorption becomes important
at $\sim 10^{11}-10^{12}$ Hz and that the emission is, of course,
substantially suppressed below this frequency.  Interestingly, Figure
\ref{plot:EIGHT} also shows that the fractional variability is also
suppressed below the self-absorption frequency.  The reason is simple:
optically thin calculations erroneously include emission from small
radii, where the flow is strongly variable.  By contrast, when optical
depth effects are included the photosphere moves out to $\sim 10-50
R_s$ (depending on frequency).  The net effect is that the fractional
variability is smaller when synchrotron self-absorption is accounted
for because the emission from small radii is excluded.

We have also calculated the polarization of the synchrotron emission
including optical depth effects (Fig. \ref{plot:NINE}). We neglect the
effects of Faraday rotation in this calculation and again consider
only thermal electrons with $T_e/T_{tot} = 1/4$. These calculations,
in contrast to the optically thin polarization calculations shown in
Figure \ref{plot:SIX}, place the observer in the equatorial plane of
the disk, thus maximizing the observed polarization.  As Figure
\ref{plot:NINE} shows, the dominant effect of self-absorption is to
substantially decrease the polarization fraction at self-absorbed
frequencies $\lsim 10^{11}$ Hz.  This is because emission from the
central optically thick impact parameters dominates the emission from
peripheral optically thin impact parameters, and optically thick
thermal emission is unpolarized. Figure \ref{plot:NINE} also shows
that the variability of the polarization fraction remains largely
unchanged when optical depth effects are taken into account. The angle
of the polarization is also unchanged, since it is still determined by
the predominantly toroidal field.

\section{Discussion}

Our results on synchrotron radiation from MHD simulations of
radiatively inefficient accretion flows can be succinctly summarized
as follows: (1) the emission is highly variable: the flux can change
by up to an order of magnitude on time-scales as short as the orbital
period near the last stable orbit ($\sim 1$ hour for Sgr A*).  (2) the
variability is stronger and more rapid at high photon frequencies
because the high frequency emission arises from closer to the black
hole; (3) The emission at self-absorbed frequencies is less variable
than the emission at optically thin frequencies.  This is in part a
consequence of point (2); optically thick emission arises further from
the hole where the variability is weaker and less rapid.  (4) At
optically thin frequencies, the synchrotron emission is linearly
polarized at the $\sim 20-50 \%$ level (unless the flow is viewed
face-on or there is significant Faraday depolarization); the
polarization vector is perpendicular to the equatorial plane of the
accretion flow and is due to the coherent toroidal magnetic field. The
magnitude of the polarization is itself significantly variable (by
factors of few). (5) For a roughly thermal electron distribution
function, the polarization fraction decreases significantly at
optically thick frequencies.

We believe that these results are reasonably robust in spite of
several significant uncertainties in our analysis.  In particular, how
we treat the electron temperature in the flow does not significantly
change these conclusions (see Fig. 5).  We have also checked that our
conclusions are unchanged if we only consider emission from gas
outside of (say) $4 R_s$ (neglecting emission from between $2-4 \
R_s$).  This is encouraging because our pseudo-Newtonian treatment of
the dynamics is particularly inaccurate at very small radii.  We
suspect that the results presented here are characteristic of models
in which the emission is dominated by gas in the accretion flow
itself.  If, however, there is preferential electron
acceleration/heating in a low density corona or in an outflow, this
could lead to variability and polarization signatures significantly
different from those presented here.  A much more careful treatment of
the electron thermodynamics (beyond MHD) is required to assess this
uncertainty. In future work, it would also be interesting to carry out
the radiative transfer including Faraday rotation and General Relativistic
photon transport.
 
\subsection{Application to Sgr A*}

We suggest that the conclusions highlighted above regarding
synchrotron emission in the MHD simulations are reasonably consistent
with observations of Sgr A* from the radio to the X-ray (summarized in
\S1).  For example, our calculations produce significant linear
polarization in the mm-IR emission, in accord with observations (Bower
et al. 2003; Genzel et al. 2003). We also find a strong decrease in
the linear polarization fraction with frequency below $\sim 100$ GHz
due to optical depth effects. It is probably not surprising that the
precise values of the observed polarization fractions are not
recovered in our calculations; we do not as yet include Faraday
depolarization, which should further decrease the polarization
fraction (Quataert \& Gruzinov 2000), nor can we easily constrain the
overall density of the flow, which determines the exact frequency at
which optical depth effects set in.

The calculations presented here predict that the polarization angle is
perpendicular to the equatorial plane of the accretion flow, and
should not change significantly with frequency.  The former prediction
is somewhat tricky to test, however, because it is not obvious what
the orientation of the accretion flow is on the sky.  The
Bardeen-Petterson effect may not be efficient for thick accretion
disks (e.g., Natarayan \& Pringle 1998), so it is unclear whether the
angular momentum of the flow at small radii is tied to that of the
hole.  If not, the orientation of the flow may be set by the angular
momentum of the stars whose winds feed the hole (see Levin \&
Beloborodov 2003 and Genzel et al. 2003a for discussions of the
stellar angular momentum).  If, however, the disk's angular momentum
is tied to the hole's, then the orientation of the flow at small radii
will depend on the angular momentum accreted by the hole over its
lifetime, which is not known.

Our calculations naturally produce factors of $\sim 10$ variability in
the IR emission on $\sim$ hour time-scales (Fig. 3), and thus may
account for some of the observed IR variability from Sgr A*.  They do
not, however, quite produce sufficient variability on time-scales as
short as is observed, $\sim 10$s of minutes (less than the orbital
period at the last stable orbit for a non-rotating black hole).  This
could be because our calculations have not been carried out for
rotating black holes or because the most rapid IR variability traces
particle acceleration, not the accretion flow dynamics.\footnote{Note
also that our spatial resolution is the poorest at the small radii
where the IR emission is produced, so we may not fully resolve
structures (e.g., turbulent eddies) that vary on $\sim 10$ minute
time-scales.}  It is worth noting that we do not see any evidence for
quasi-periodic oscillations in our calculations, though we have not
included Doppler boosting that could modulate the emission on the
orbital time-scale (e.g., Melia et al. 2001).

\begin{figure}
%\begin{center}
%\includegraphics[height=6.6in, width=5.1in, angle=90]{FIGUREEIGHT}
\hspace*{-.1cm}
\centerline{
\psfig{file=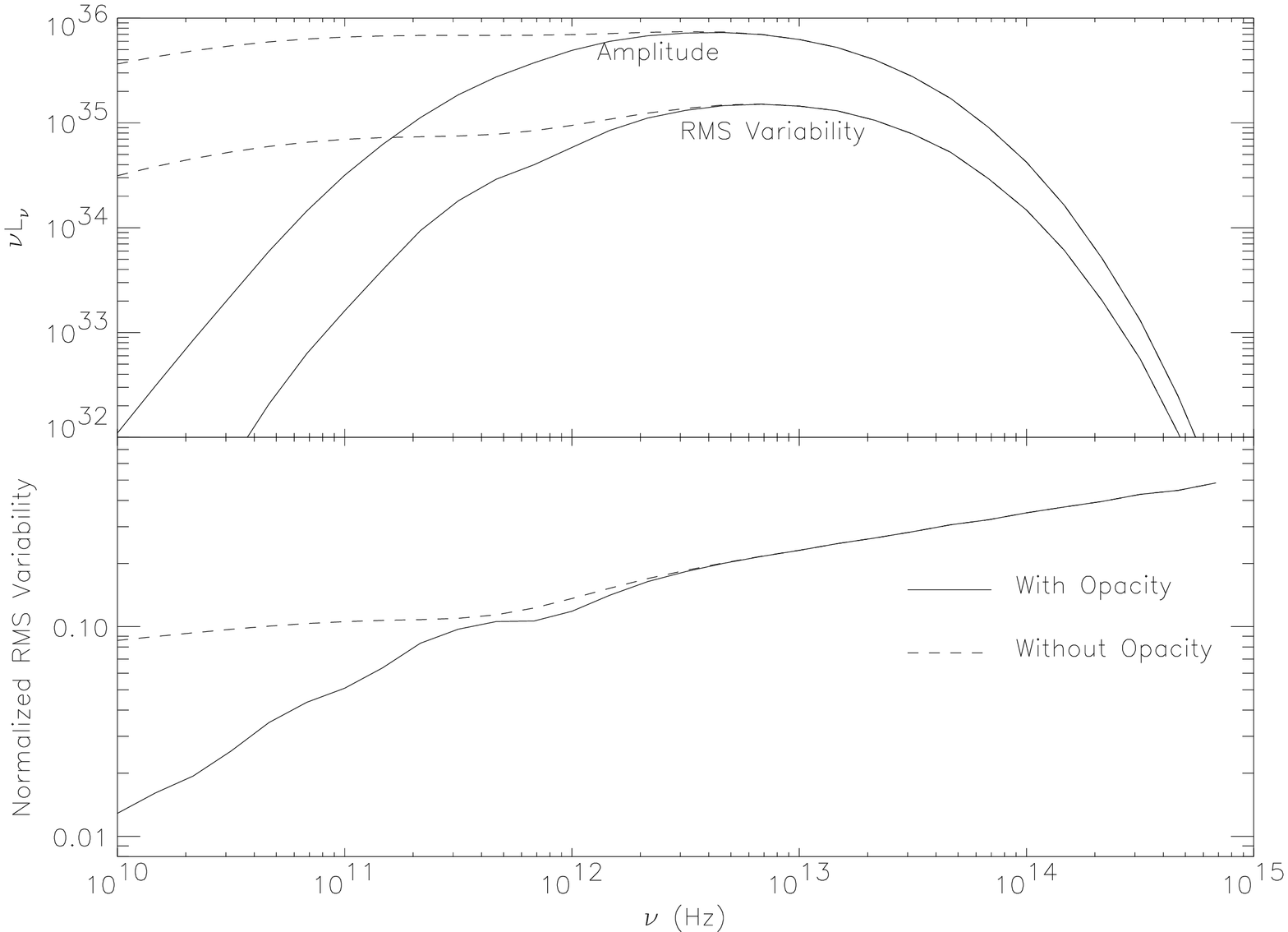,angle=90,width=7.3in}}
%\plotone{FIGUREEIGHT.eps}
%\end{center}
\caption{The top panel displays the mean ($\nu L\nu$) and the RMS
variability of the emission including (solid line) and excluding
(dashed line) optical depth effects. The bottom panel shows the
variability as a fraction of the mean. Note that the RMS shown here
yields a quantitative measure of variability that is systematically
smaller (by a factor of few) than the Fourier analysis shown in Figure
5 (see text for more discussion).}
\label{plot:EIGHT}
\end{figure}

\begin{figure}
%\begin{center}
%\includegraphics[height=6.6in, width=5.1in, angle=90]{FIGUREEIGHT}
\hspace*{-.1cm}
\centerline{
\psfig{file=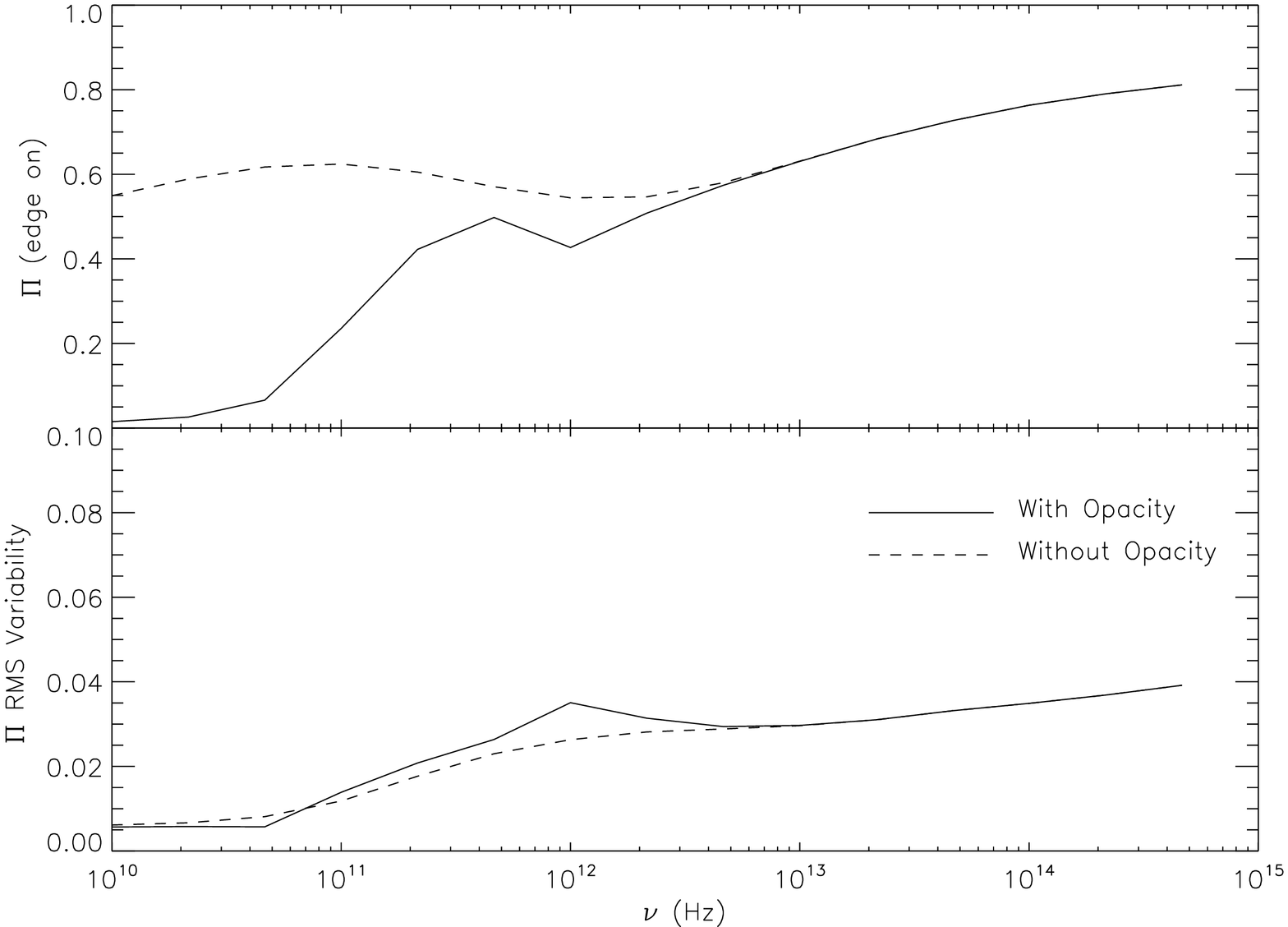,angle=90,width=7.3in}}
%\plotone{FIGUREEIGHT.eps}
%\end{center}
\caption{The top panel shows the average polarization fraction ($\Pi$)
of the emission including (solid line) and excluding (dashed line)
optical depth effects (averaged over all 32 time-slices for which we
have spatial data). The bottom panel shows the RMS variability of the
polarization fraction. These calculations assume purely thermal
electrons with $T_e/T_{tot} = 1/4$ and that the disk is observed edge
on.}
\label{plot:NINE}
\end{figure}

More generally, our calculations produce variability that increases in
amplitude and decreases in time-scale with increasing frequency (Figs. 3
\& 5), as is also observed from Sgr A*.  In particular, the
suppression of variability at optically thick frequencies
(Fig. \ref{plot:EIGHT}) may contribute significantly to the
differences in the observed variability in the radio, mm, and IR (it
is also possible that this is due to a change in the dynamical
component responsible for the emission; e.g., a jet becoming important
at lower frequencies as in Yuan et al. 2002).  In addition, although
we have not calculated X-ray emission, the large amplitude, hour
time-scale variability we see from gas close to the black hole is
somewhat reminiscent of the X-ray flares observed by {\it Chandra} and
{\it XMM}. {Moreover, since some of the X-ray flares could be produced
by synchrotron emission (e.g., YQN04), it is plausible to suppose that
the synchrotron variability we calculate here could extend to higher
frequencies as well.  Finally, it is worth reiterating that that the
calculations presented here are also likely lower limits to the
variability at $\nu \gsim 10^{12}$ Hz because we do not include
variations in the non-thermal electron population or effects from
synchrotron self-Compton emission.}  Such transient particle
acceleration appears to be required to explain the very large
amplitude X-ray flares (Markoff et al. 2001; YQN04).

%\footnote{Note that on
%very short time-scales our results could be somewhat affected by
%finite spatial resolution, i.e., because we cannot resolve structures
%that vary on $\sim$ 10s minutes.  We do not, however, think that this
%can account for the lack of power on $\sim 10$ minute time-scales
%because a power spectrum of the variability shows a monotonic decrease
%from $\sim$ hour times-scales to $\sim 10$ minute time-scales, with no
%evidence for artificial/numerical suppression of power.}  

Our calculations predict that variability at different frequencies
should be strongly correlated so long as both frequencies are
optically thin (see Fig. 3).  We also predict that the time delay
between the emission at different optically thin frequencies should be
quite small, $\lsim$ hours.  These predictions may be testable by
correlating sub-mm and IR variability from Sgr A*.  Note, however,
that strong temporal or spatial variations in electron acceleration
could modify these predictions.  Our results also suggest that
emission at optically thick frequencies should not be as well
correlated (though we have not shown this explicitly).  The reason is
that different frequencies then probe different radii; since the
turbulence at one radius need not be well correlated with the
turbulence at another radius, the same follows for variations in the
synchrotron emission.
 
In spite of a few shortcomings, the general agreement between the
properties of synchrotron emission in our calculations and
observations of the Galactic Center is encouraging.  It supports a
model in which much of the high frequency emission from Sgr A* is
generated by a turbulent magnetized accretion flow close to the black
hole, and encourages more refined calculations of emission from
numerical simulations of RIAFs.
 
\acknowledgments We thank Eric Agol, Geoff Bower, and Jim Stone for
useful discussions, and the referee for very useful comments and
suggestions. JEG also wishes to thank Evan Levine and Erik Rosolowsky
for their copious and insightful assistance. JEG and EQ were supported
in part by NSF grant AST 0206006, NASA Grant NAG5-12043, an Alfred
P. Sloan Fellowship, and the David and Lucile Packard Foundation. JEG
was supported in part by NSF grant AST 04-06987. IVI
was supported by the U.S. Department of Energy (DOE) Office of
Inertial Confinement Fusion under Cooperative Agreement
No. DE-FC03-92SF19460, the University of Rochester, the New York State
Energy Research and Development Authority.
 
%\newpage
%\vspace{-1in}

%\newpage

\clearpage

\begin{figurehere}
\vspace*{1.2in}
\centerline{
\hspace{-0.5cm}
\psfig{file=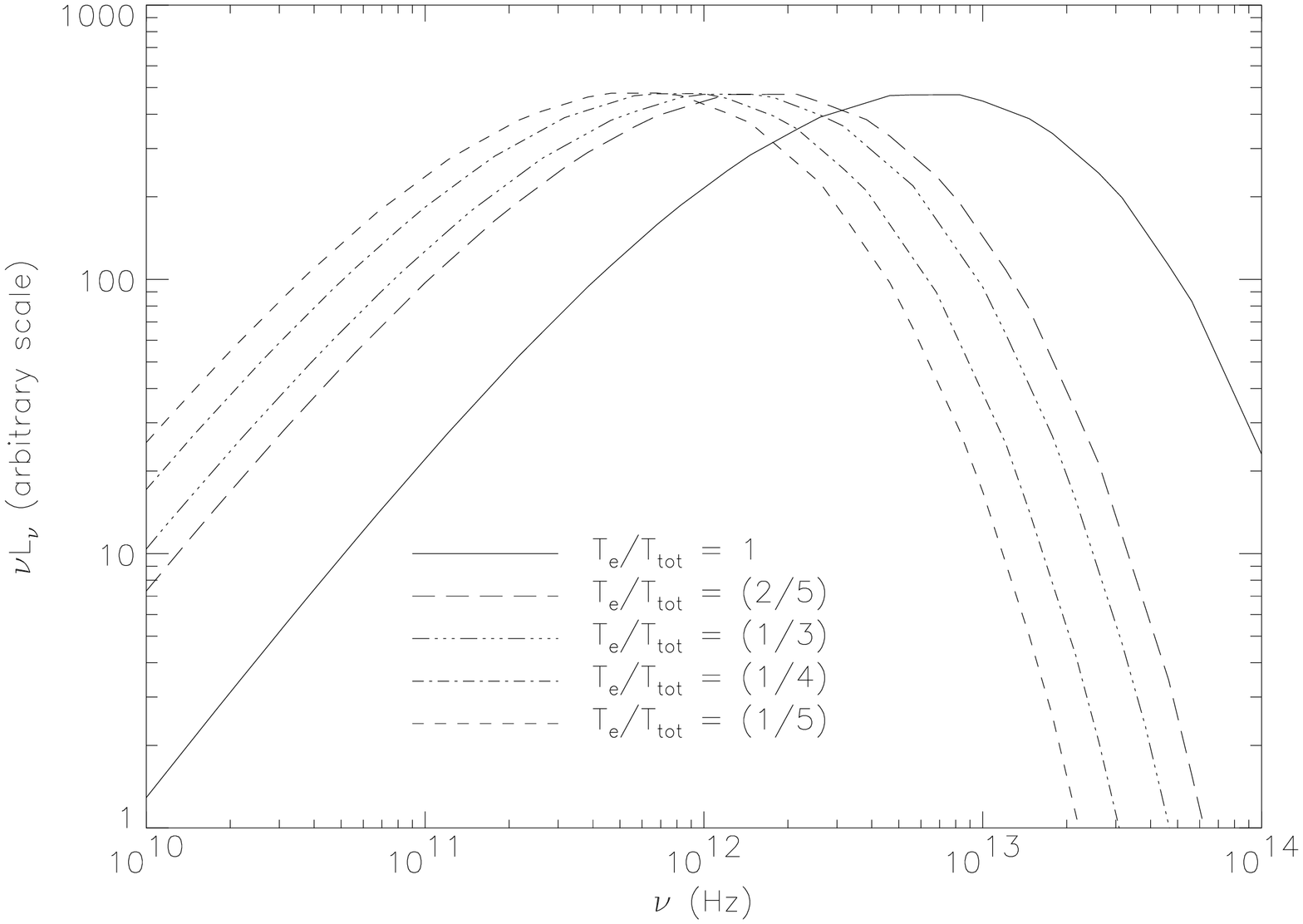,angle=90,width=7.5in}}
%\plotone{FIGUREEIGHT.eps}
%\plotone{FIGUREEIGHT.eps}
%\end{center}
\caption{$\nu L_{\nu}$ from an arbitrary volume of gas with $n_e = q \times 10^6$ cm$^{-3}$, $B= \sqrt{q} \times 70 \ G$, and $T_e = 2 \times 10^{11} \left( T_e/T_{tot}\right) {\rm K}$, for various values of $T_e/T_{tot}$ and $q$; note that $B \propto \sqrt{q}$ is equivalent to $\beta = {\rm const.}$  In each case the density parameter $q$ has been adjusted so that the total synchrotron power is roughly the same: $q \approx 2.49, 6.23, 7.56, 10, \ \& \ 12.5$ for $T_e/T_{tot} =1, 2/5, 1/3, 1/4, \ \& \ 1/5$.}
\label{plot:ONE}
\end{figurehere}
%\clearpage

\clearpage

\clearpage

%\clearpage
%\begin{figure}
%\begin{center}
%\includegraphics[height=6.6in, width=5.1in, angle=90]{FIGURETWOA}
%\end{center}
%\caption{Two A.}
%\label{plot:TWOA}
%\end{figure}

\clearpage

\clearpage

%\clearpage
%\begin{figure}
%\begin{center}
%\includegraphics[height=6.6in, width=5.1in, angle=90]{FIGUREFIVE}
%\end{center}
%\caption{Power spectrum of data set 2. Note that this power spectrum is qualitatively very similar to higher photon frequency spectra, into the infrared.}
%\label{plot:FIVE}
%\end{figure}

\clearpage

\clearpage

\clearpage

\end{document}